\documentclass[twocolumn]{aa}
\usepackage[]{natbib}
\usepackage{graphicx,color}
\usepackage{txfonts}
\usepackage[colorlinks=true, citecolor=blue]{hyperref}
\usepackage[table]{xcolor}
\usepackage{multirow} %multiple row in tables
\usepackage{boldline} %bold lines in tables
\usepackage{adjustbox}
\usepackage{float}
\usepackage{booktabs}

\newcommand{\orcid}[1]{\href{https://orcid.org/#1}{\includegraphics[width=10pt]{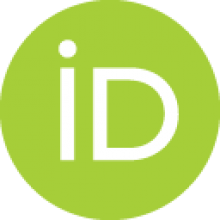}}}

\begin{document} 
\title{A Newly Discovered VVVX Star Cluster Surfing the Galactic Disk}

\author{
Elisa R. Garro\inst{1}\thanks{Corresponding author; elisaritagarro1@gmail.com}\orcid{0000-0002-4014-1591}
\and 
Jos\'e G. Fern\'andez-Trincado\inst{2}\thanks{Corresponding author; jose.fernandez@ubo.cl}\orcid{0000-0003-3526-5052}
\and Casmir O. Obasi\inst{3}\orcid{0000-0003-3526-5052}
\and Dante Minniti\inst{3,4}\orcid{0000-0002-7064-099X}
\and Matías Gómez\inst{3}\orcid{0000-0002-4430-9427}
\and Roberto K. Saito\inst{5}\orcid{0000-0001-6878-8648}
\and Javier Alonso-García\inst{6}\orcid{0000-0003-3496-3772}
\and Alonso Luna\inst{1}\orcid{0000-0001-5971-8058}
\and Valentin D. Ivanov\inst{7}\orcid{0000-0002-5963-1283}
\and Maria Gabriela Navarro\inst{9}\orcid{0000-0002-1860-2304}
\and Sara Federle\inst{3}\orcid{0009-0001-8940-6124}
\and Bernardo P. L. Pereira\inst{10}\orcid{0000-0002-7552-3063}
\and Beatriz Barbuy\inst{10}\orcid{0000-0001-9264-4417}
\and Devika Bhadrakumar Sindhu\inst{6}
\and Bruno Dias\inst{3}\orcid{0000-0003-4254-7111}
\and Tali~Palma\inst{7,8}\orcid{0000-0002-0732-2737}
\and Ilaria Petralia\inst{3}\orcid{0009-0000-2403-9442}
\and Sasi Saroon\inst{3}\orcid{0000-0002-2789-0934}
}

\authorrunning{Garro et al.} 
	
\institute{
ESO - European Southern Observatory, Alonso de Cordova 3107, Vitacura, Santiago, Chile
\and 
Centro de Investigaci\'on en Astronom\'ia, Universidad Bernardo O’Higgins, Avenida Viel 1497, Santiago, Chile
\and
Instituto de Astrofísica, Depto. de Físicas y Astronómia, Facultad de Ciencias Exactas, Universidad Andres Bello, Av. Fernandez Concha 700, Las Condes, Santiago, Chile
   \and
 Vatican Observatory, Vatican City State, V-00120, Italy
 \and
Departamento de Física, Universidade Federal de Santa Catarina, Trindade 88040-900, Florianópolis, SC, Brazil
\and
Centro de Astronomía (CITEVA), Universidad de Antofagasta, Av. Angamos 601, Antofagasta, Chile
\and
ESO - European Southern Observatory, Karl-Schwarzschild-Strasse 2, 85748 Garching bei München, Germany
\and
INAF - Osservatorio Astronomico di Roma, Via di Frascati, 33, 00078, Monte Porzio Catone, Italy
\and 
Universidade de Sao Paulo, IAG, Departamento de Astronomia, 05508-090 Sao Paulo, Brazil
\and 
Observatorio Astronómico, Universidad Nacional de Córdoba, Laprida 854, X5000BGR, Córdoba, Argentina
\and 
Consejo Nacional de Investigaciones Científicas y Técnicas de la República (CONICET), Godoy Cruz 2290, C1425FQB, CABA,
Argentina
}
		
\date{Received 5/07/2026; Accepted 20/09/2026}

	\abstract
	% context heading (optional)
	{
We report the discovery of Garro 04, a previously unidentified stellar cluster toward the Galactic disc, uncovered as part of the ESO/KMOS VVVX-GalCen Spectroscopic Survey. This object was serendipitously confirmed as a stellar overdensity in the VISTA Variables in the V\'ia L\'actea eXtended (VVVX) survey footprint and subsequently confirmed through medium-resolution $K$-band spectroscopy obtained with the K-band Multi Object Spectrograph (KMOS) at the ESO Very Large Telescope. The combined photometric and spectroscopic analysis reveals a coherent stellar population with common radial velocities and chemical properties, supporting its classification as a genuine star cluster rather than a statistical fluctuation of the disk field. We derive the fundamental properties of Garro 04, including its structural parameters, distance, kinematics, and compute its orbital properties, finding that its motion remains confined to the Galactic disk region. Finally, our analysis suggests that Garro~04 is likely an old open cluster, or an object straddling the boundary between the canonical globular and open cluster classifications, with an estimated Age $ = 5 \pm 2$ Gyr and a metallicity of $\mathrm{[Fe/H]} = -0.5 \pm 0.2$, finding an excellent agreement between the photometric and spectroscopic determinations.
}

	\keywords{Galaxy: disk; Galaxy: open clusters and associations: individual (Garro~04); Techniques: photometric, spectroscopic, radial velocities; Surveys}
	\maketitle
	
%%%%% INTRODUCTION %%%%%%%
\section{Introduction} \label{sec:intro}

The inner regions of the Milky Way, particularly the Galactic plane, constitute one of the most challenging yet rewarding environments for stellar astrophysics. Characterized by extreme stellar densities, intense star formation activity, and pervasive interstellar extinction from dust lanes, this region has historically remained largely opaque to optical observations. Bound stellar systems serve as critical tracers of the Galaxy’s chemical enrichment history, dynamical evolution, and star formation processes over cosmic time \citep{2025A&A...696A.179F, 2025arXiv251213590S}. Understanding their properties is essential for reconstructing the assembly and secular evolution of the Milky Way’s central regions, yet the obscuration has long limited comprehensive studies of these systems.

The advent of large-scale near-infrared photometric surveys has dramatically altered this landscape. The VISTA Variables in the Via Láctea (VVV, \citealt{2010NewA...15..433M, 2012A&A...537A.107S}) survey and its eXtension (VVVX; \citealt{2024A&A...689A.148S}) have delivered deep, high-resolution multi-epoch imaging across the Galactic bulge and inner disk, penetrating the heavy dust extinction that obscures optical wavelengths. By exploiting near-infrared passbands, these campaigns have enabled the detection and characterization of stellar overdensities that were previously undetectable \citep{Minniti_2017,2002A&A...394L...1I}. As a result, VVV/VVVX data have facilitated the systematic identification of hundreds of previously unknown open and globular cluster (GC) candidates, many of which are young or intermediate-age\footnote{When referring to intermediate-age objects, we mean star clusters that appear too young to be classified as GCs and too old to be considered typical OCs. In general, GCs are characterized by ages older than $\sim 10$ Gyr, whereas most OCs are younger than $\sim 1$--2 Gyr. Nevertheless, a few notable exceptions exist, such as NGC 6791, which has an age of approximately 8 Gyr.} systems embedded within dense molecular clouds or lying behind significant columns of interstellar material (e.g. \citealt{2024A&A...687A.214G} and references therein).

These discoveries have substantially expanded the known population of Milky Way stellar clusters, revealing a diverse sample that includes metal-rich and metal-poor objects, as well as candidates spanning a wide range of ages and masses \citep{garro22c, 2021A&A...652A.129M, 2020RNAAS...4..218M, 2020MNRAS.499.3522B}. The catalogs derived from VVV/VVVX have provided new insights into the spatial distribution, luminosity functions, and orbital properties of clusters within the complex gravitational potential of the Galactic bar and bulge. Moreover, proper-motion (PM) cleaning and color-magnitude diagram (CMD) analysis have allowed for the separation of cluster members from the overwhelming field-star contamination, yielding robust membership lists and preliminary physical parameters for numerous systems \citep{2024A&A...687A.214G, 2024A&A...688L...3G, 2022A&A...658A.120G, garro22b}. The power of PM-based searches has been further boosted by the {\it Gaia} mission \citep[DR2 and DR3;][]{2018A&A...616A...1G, 2023A&A...674A...1G}, whose homogeneous, high-precision astrometry over the entire sky has enabled the systematic, blind identification of new clusters through PM and parallax clustering; a striking example is provided by \citet{2022A&A...661A.118C}, who reported 628 new open clusters from a search of {\it Gaia} EDR3 with the OCfinder method. Nevertheless, the census of Milky Way GCs remains incomplete \citep{2024NewAR..9901696C, Minniti_2017, 2005A&A...442..195I}, with many faint or highly reddened objects still awaiting confirmation and detailed characterization.

Motivated by these gaps, continued exploration of the VVV/VVVX database keeps revealing additional hidden stellar systems. Advanced selection techniques that combine photometric variability with PM and statistical field decontamination have proven particularly effective in identifying genuine cluster candidates against the dense stellar background \citep{2025A&A...695A.235O, 2024A&A...689A.115S, garro24}. Such efforts not only refine our understanding of the bulge and disk's stellar content but also highlight the surveys' ongoing potential to reveal structures that challenge or refine current models of Galactic evolution. The identification of these objects underscores the transformative role of infrared astronomy in modern Galactic studies.

In this work, we report the discovery of a new stellar cluster in the Galactic disk, Garro~04 (Section~\ref{sec:discovery}). This previously unrecognized system is identified as a statistically significant stellar overdensity with coherent proper motions and a well-defined color-magnitude sequence (Sections~\ref{sec:validation} and~\ref{sec:physicalparm}), adding a new member to the growing family of infrared-discovered clusters. We further present spectroscopic data (Section~\ref{sec:kmosdata}) and derive radial velocities and orbital parameters (Section~\ref{sec:orbits}). The methodology adopted to derive the spectroscopic metallicity is described in Section~\ref{sec:specmet}. Together, these results provide fresh constraints on star formation and dynamical processes in the inner Milky Way (Section \ref{sec:Sum}).

\section{Discovery and observational datasets}
\label{sec:discovery}
\begin{figure}[htpb]
    \centering
    \includegraphics[width=\linewidth]{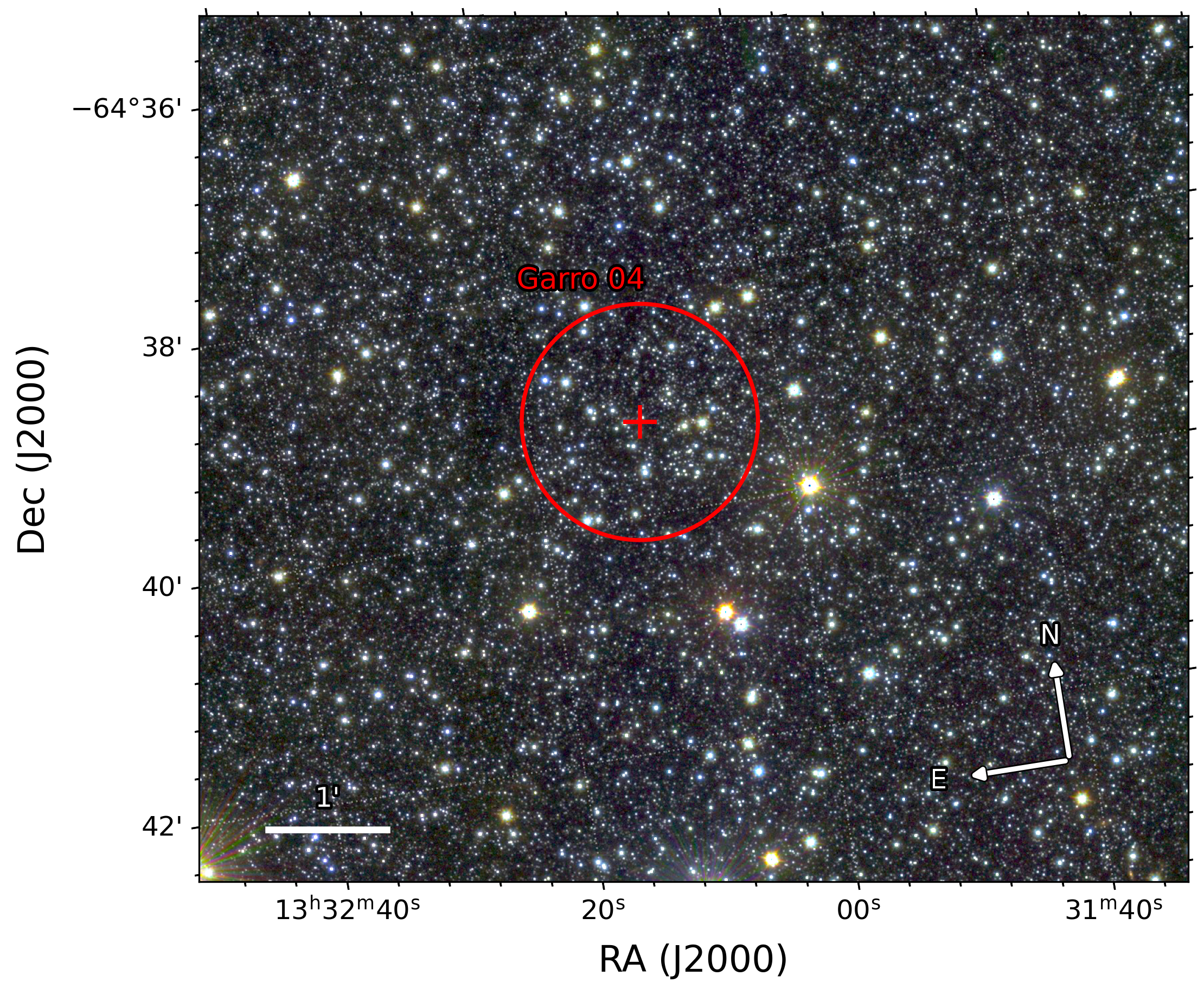}
\caption{VVVX $JHK_s$ colour-composite image of the Garro\,04 field
(with a field of view of $8\farcm4\times7\farcm3$), centred on RA $=13^{\rm h}32^{\rm m}07\fs5$,
Dec $=-64\degr39\arcmin30\farcs5$ (J2000). The red cross marks the new cluster centre
(RA $=13^{\rm h}32^{\rm m}11\fs33$, Dec $=-64\degr39\arcmin12\farcs60$, J2000) and
the red circle is $1\arcmin$ in radius. North and East are
indicated by the arrows; the white scale bar corresponds to 1\,arcmin.}
    \label{figure1}
\end{figure}

We report the discovery of a new star cluster candidate (Fig. \ref{figure1}) -- hereafter Garro~04 -- located at  $R.A. = 13^{\rm h}32^{\rm m}07.5^{\rm s}$, $Dec = -64\degr39\arcmin30.5\arcsec$ (J2000;  equivalent to $\alpha = 203.03\degr$, $\delta = -64.66\degr$ and Galactic coordinates $l=307.270\degr$ and $b=-2.130\degr$), identified as a conspicuous overdensity of red clump stars through direct visual inspection of the VVVX tile \texttt{d009}, analogously to our previous cluster discoveries in the VVV/VVVX footprint. The analysis follows the same  methodology previously adopted for Garro~01 and Garro~03 \citep{garro20,garro24}, two cluster candidates in angular projected proximity of $\sim 3.2\degr$ to Garro~04.\\

We made use of a combination of optical and near-infrared (NIR) public surveys to confirm the existence of Garro~04 and characterize its nature. The VVVX data were retrieved from the ESO Science Archive Facility\footnote{\url{http://archive.eso.org}} (programme IDs 179.B-2002 and 198.B-2004), the {\it Gaia} Data Release~3 (DR3; \citealt{gaia23}) from the ESA \textit{Gaia} Archive\footnote{\url{https://gea.esac.esa.int/archive/}}, and the Two Micron All Sky Survey (2MASS; \citealt{skrutskie06}) data from the VizieR catalogue access tool at CDS\footnote{\url{https://vizier.cds.unistra.fr}}. In particular, we first built a combined NIR and optical catalog by cross-matching the VVVX and {\it Gaia} DR3 sources within a matching radius of $0.5\arcsec$. The VVVX is a public ESO survey carried out with the VISTA Infrared CAMera (VIRCAM) mounted on the 4.1\,m wide-field Visible and Infrared Survey Telescope for Astronomy (VISTA; \citealt{2010Msngr.139....2E}). PSF photometry was extracted following the procedure described in \citet{2018A&A...619A...4A,2026A&A...706A.301A}. Astrometry was calibrated to the {\it Gaia} DR3 reference frame \citep{gaia18a}, while photometric zero-points were tied to the VISTA magnitude system \citep{Gonzalez_Fernandez2018} via the 2MASS dataset, using a globally optimized frame-by-frame model that incorporates an illumination correction. As a further check, visual inspection of the {\it Gaia}~DR3 images reveals a clear round stellar overdensity at the same coordinates, as shown in Fig. \ref{fig:VPM_dec_and_new_centre}. Exploiting {\it Gaia}'s precise astrometry and PMs, we applied two quality filters to the catalog: we retained only sources with parallax 
$\varpi > 0.5$\,mas (corresponding to $D < 3$\,kpc; \citealt{BailerJones2018}) to exclude 
nearby foreground contamination, and with \texttt{RUWE} $< 1.4$ to ensure reliable astrometric solutions \citep{Fabricius2021}. The {\it Gaia} PMs were then used to kinematically separate probable cluster members from field stars and derive the PM-membership probability for each star, as explained in Section \ref{sec:validation}. Moreover, since the VVVX images are saturated for $K_s<11$, we completed the upper part of the CMDs using a combination of {\it Gaia} DR3 and 2MASS catalogs. We treated the two NIR datasets separately, but we scaled them to the same photometric system \citep{Gonzalez_Fernandez2018}, and applied correction offsets of $\Delta K_s < 0.009$ mag and $\Delta J < 0.051$ mag. 

\section{Decontamination procedure and cluster membership}
\label{sec:validation}
From Fig.~\ref{figure1} alone, it is not clear whether a circular overdensity of stars is present. Therefore, to validate the existence of Garro~04 and rule out the possibility that the detected overdensity is merely a chance grouping of stars, a disk fluctuation, or an asterism, we follow the statistical procedure described in \citet{garro22c}. A concise summary of the main steps is given below.

We first employ the kernel density estimation (KDE) technique and overlay iso-density contours to verify the presence of a central stellar overdensity. We expected to manifest as a centrally concentrated density peak decreasing smoothly outward, and to place preliminary constraints on the cluster extent. We retained stars within $\sim 3'$ for subsequent analysis.

\begin{figure*}[htpb]
\centering
\includegraphics[width=0.3\textwidth]{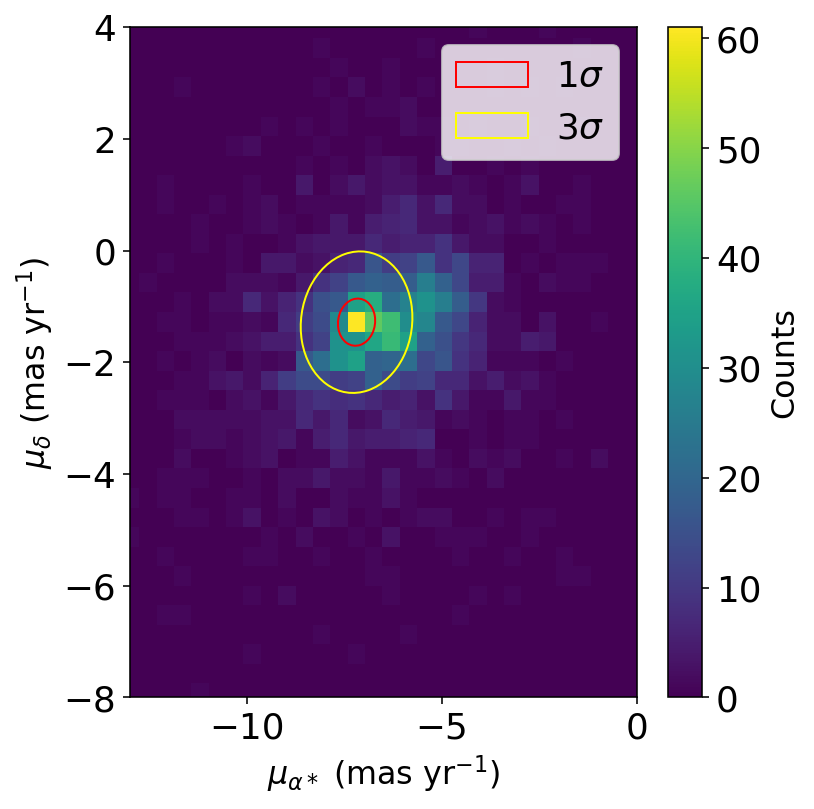}
\includegraphics[width=0.62\textwidth]{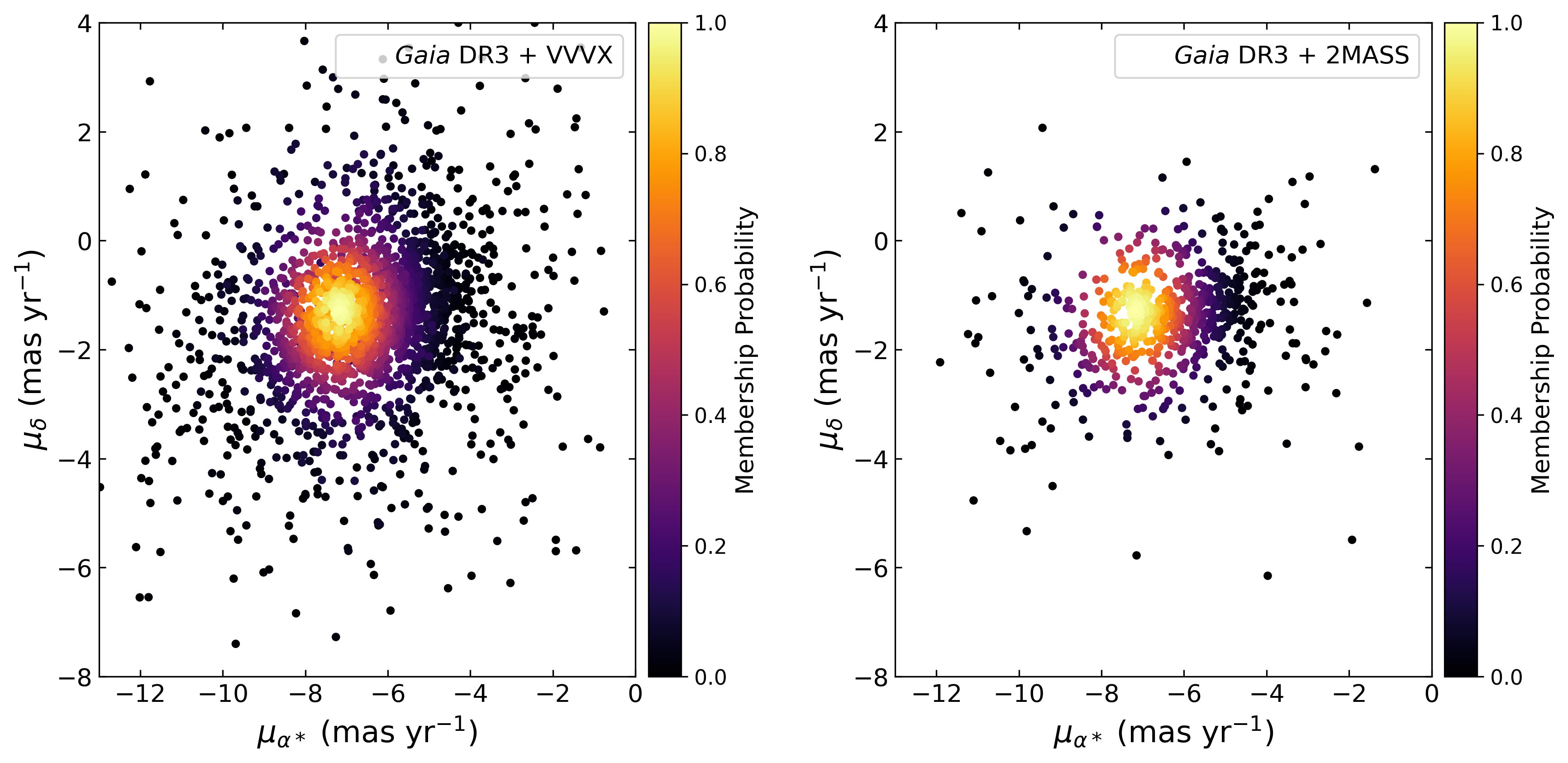}
\includegraphics[width=0.4\textwidth]{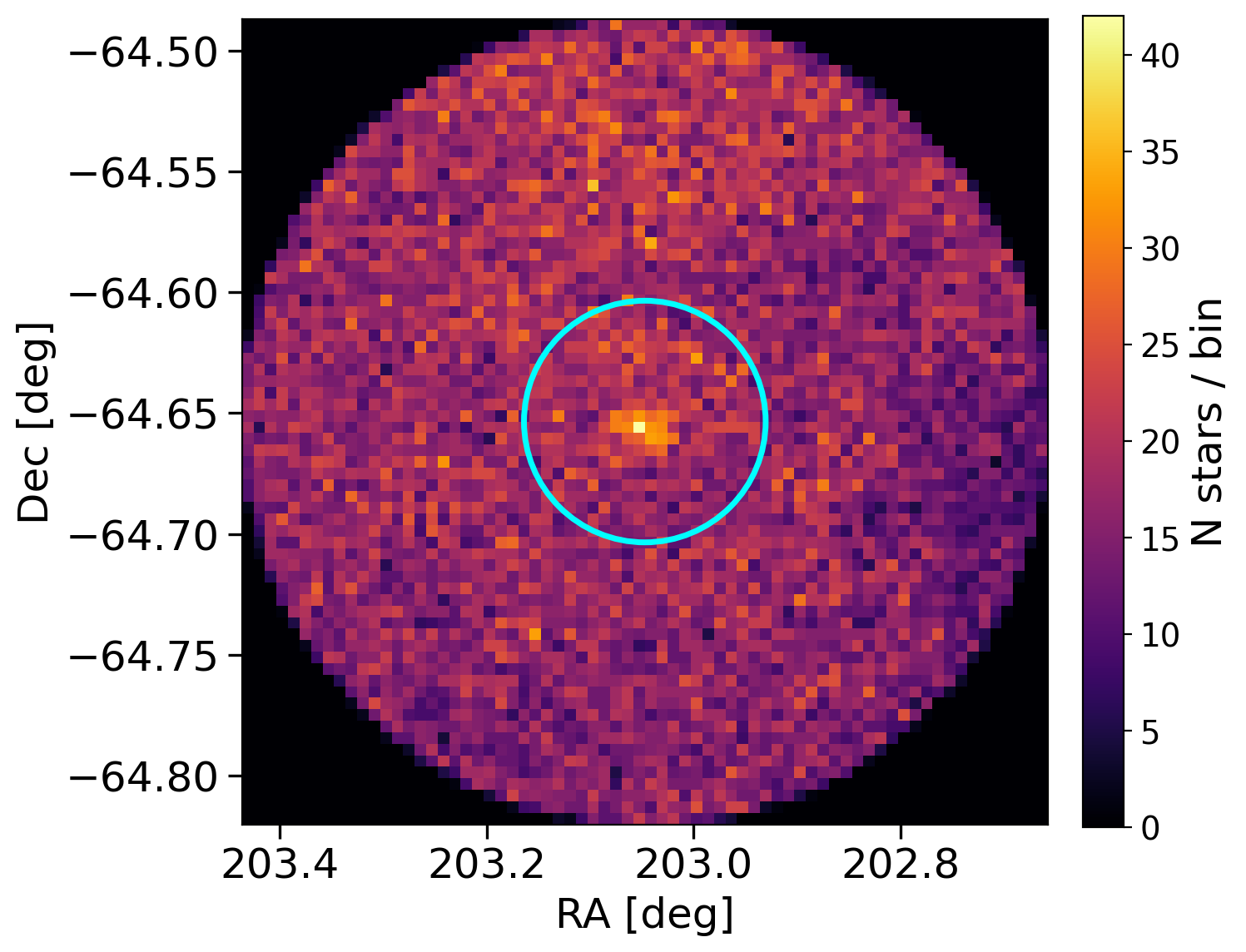}
\includegraphics[width=0.4\textwidth]{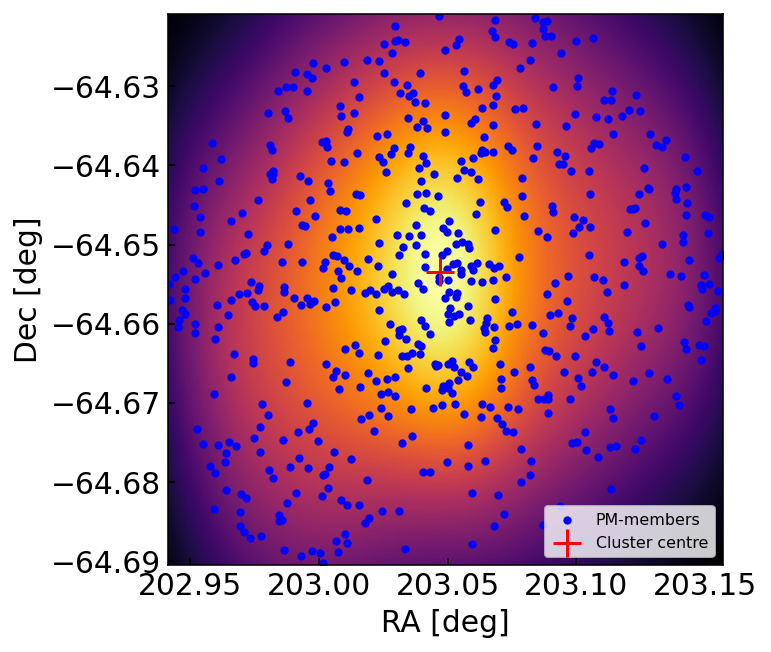}
\includegraphics[width=0.4\textwidth]{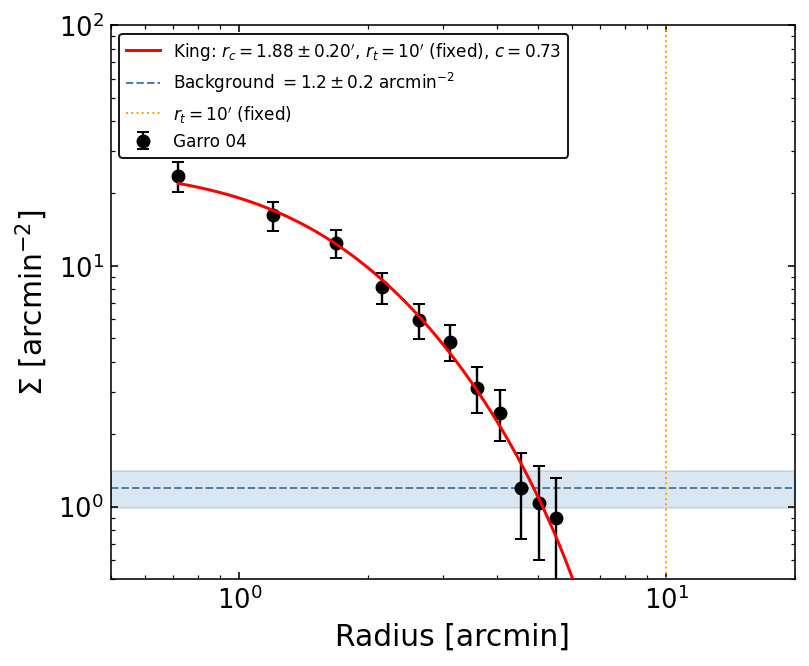}
\caption{{\bf Upper panels.} Vector proper motion (VPM) diagram for stars in the Garro~04 field, shown as a 2D density map (bin counts indicated by the colour bar). The strongest density peak (the yellower bin) marks the cluster's mean proper motion and the red and yellow ellipses denote the $1\sigma$ and $3\sigma$ contours of the proper motion distribution around the peak.\textit{Middle and right panels:} 
VPM for the  VVVX+{\it Gaia}~DR3 (\textit{left panel}) and 2MASS+{\it Gaia}~DR3 (\textit{right panel}) samples, respectively, with stars color-coded by PM membership probability as indicated by the color bar. {\bf Middle panels.} \textit{Left panel:} \textit{Gaia} DR3 stellar-density map within $10\arcmin$ of the cluster centre, without applying any PM or photometric cut. The cyan circle marks a $3\arcmin$ radius around the newly identified cluster centre. A clear central overdensity is recovered, independently confirming the reality of the cluster. \textit{Right panel:} New center determination and projected density profile. As indicated in the legend, the blue points represent the PM selected cluster members. The color scale reflects the KDE stellar surface  density, with yellow indicating the highest density regions, decreasing through orange and blue to black at the lowest density levels.  {\bf Bottom panel}. The black points show the background-corrected radial density profile of Garro~04, with error bars including the Poisson uncertainty of the background. The red solid line
shows the best-fit \citet{1962AJ.....67..471K} model, yielding a core radius $r_c = 1.88 \pm 0.20$\,arcmin for a fixed tidal radius $r_t = 10'$ (orange dotted line). The blue
dashed line marks the adopted background level, and the shaded band its Poisson uncertainty.}
\label{fig:VPM_dec_and_new_centre}
\end{figure*}

Given the severe dust extinction and high stellar crowding characteristic of this line of sight, we rely on PMs to kinematically disentangle cluster members from the field population. To this end, we construct the vector point diagram (VPD). As shown in Fig.~\ref{fig:VPM_dec_and_new_centre}, the cluster and field populations display broadly similar PM distributions. Testing different histogram binnings -- particularly along the R.A. direction -- reveals a clearer separation, with the highest peak corresponding to the cluster. To calculate the mean cluster PM, we employed two methods: a $\sigma$-clipping iterative rejection scheme and a two-dimensional histogram analysis, identifying the strongest peak in the PM space.  This yields a mean cluster PM of $\mu_{\alpha}^{\ast} = -7.08 \pm 0.52$ mas~yr$^{-1}$ and $\mu_{\delta} = -1.36 \pm 0.48$ mas~yr$^{-1}$, selecting only stars within $1\sigma$. Using the derived mean cluster PM, we compute a membership probability for each star in the catalog following the formalism of \citet{1971A&A....14..226S}, as detailed in \citet{2025A&A...695A.235O}. Only stars with membership probability $p > 0.5$ are retained in the final catalog. The same procedure is  then applied to the {\it Gaia}~DR3+2MASS catalog, from which we likewise select stars with $p > 0.5$. Within the $\sim3'$ radial extent of the adopted catalog, the PM selection retains only 613 of the 2145 stars ($\approx29\%$) passing the quality cuts as likely cluster members, rejecting the remaining $\approx71\%$ as field contamination. The fraction of likely members increases monotonically toward the centre, from $\approx29\%$ over the full field to $\approx41\%$ within the central $0.5'$, confirming the central concentration of the cluster.
However, as shown in Fig. \ref{fig:cmd_varyingagemet}, some degree of contamination remains, particularly in the blue region of the CMD. These stars are likely foreground disk field stars whose kinematic properties are comparable to those of the cluster population, allowing them to survive the decontamination process. Nonethless, we cannot rule out that some of them are genuine cluster members and, in that case, blue straggler candidates -- although establishing their nature lies beyond the scope of the present analysis. \\

Using the cleaned membership catalog, we re-determine the center of Garro~04 following the procedure described in \citet{garro22c}. Since our sample predominantly comprises evolved stars, we adopt a 2-dimensional Gaussian KDE to locate the  position of maximum spatial stellar density (Fig. \ref{fig:VPM_dec_and_new_centre}), rather than relying on  the surface brightness maximum, which could be biased by the presence of a few luminous stars.
We evaluated a Gaussian kernel on a regular grid covering the full  extent of the member sample, with the bandwidth set to a fraction of  the total spatial extent of the data. To assess the robustness of the  result, we performed two independent stability tests: varying the  bandwidth scaling parameter over five values ($\ell = 0.5, 1.0, 1.3, 1.5, 2.0$) with fixed resolution, and varying the grid resolution over five values (200, 300, 400, 500, 600 points) with fixed bandwidth. We obtain $\alpha_{J2000} = 203.0472^{\circ}$  and $\delta_{J2000} = -64.6535^{\circ}$ -- with an uncertainty of $\pm 0.1'$ in each direction. Our KDE center lies $0.5'$ from the previously adopted position ($\alpha_{J2000} = 203.03^{\circ}$, $\delta_{J2000} = -64.66^{\circ}$), an offset of $\sim 3.6\sigma$ relative to our bootstrap centroiding uncertainty but well within the cluster core radius, reflecting the improved astrometric membership sample rather than a change in the cluster's location.\\

We constructed the radial density profile (RDP) of Garro~04 (Fig.~\ref{fig:VPM_dec_and_new_centre}) by counting PM-members in concentric annuli centered on the new cluster center, dividing by the corresponding ring area to obtain the stellar surface density $\Sigma(r)$ in arcmin$^{-2}$, and subtracting a background level of $\Sigma_\mathrm{bg} = 1.2 \pm 0.2$\,arcmin$^{-2}$ estimated from the outermost bins, where the uncertainty corresponds to the Poisson noise of the field counts (shown as the shaded band in Fig.~\ref{fig:VPM_dec_and_new_centre}). This background uncertainty is propagated into the density error bars and into the fit. We fitted the background-corrected RDP with
a \citet{1962AJ.....67..471K} model. Since the catalog extends only to $\sim$6.6\,arcmin from the cluster center, $r_t$ cannot be constrained from the data alone. We therefore fixed $r_t = 10'$, a value typical of old open clusters (OCs) at comparable Galactocentric distances \citep[e.g.][]{2008A&A...477..165P}, and fitted only $\Sigma_0$ and $r_c$ using a non-linear least-squares minimization weighted by the density uncertainties. The best-fit parameters are $\Sigma_0 = 39.3 \pm 3.6$\,arcmin$^{-2}$ and $r_c = 1.88 \pm 0.20$\,arcmin,
corresponding to a physical core radius of $r_c = 5.5 \pm 0.6$\,pc at the heliocentric distance of $d_\odot = 10.07$\,kpc (see Section~\ref{sec:physicalparm}). The fixed tidal radius corresponds to a physical extent of $r_t \approx 29.3$\,pc. The concentration parameter $c = \log(r_t/r_c) \approx 0.73$
is consistent with a moderately concentrated, tidally evolved old OC. We further verified that varying the background between 0.5 and 1.2\,arcmin$^{-2}$ changes $r_c$ by
less than $1.2\sigma$, confirming the robustness of the structural parameters against the background uncertainty.

\begin{table}[htpb]
%\onecolumn
\centering 
\renewcommand{\arraystretch}{1.1}
\caption{Garro~04 physical, kinematic, orbital and structural parameters.}
\begin{tabular}{ll}
\hline\hline
        Parameters        &  Values \\
\hline
        RA  & 13:32:11.33 \\
        DEC & -64:39:12.60 \\
        $l$ & 307.2774 [deg]\\
        $b$ & -2.1261 [deg]\\
        $A_{Ks}$ & $0.33\pm 0.04$ [mag]\\
        $A_{G}$ & $2.40\pm 0.02$ [mag]\\
        $(m-M)_0$ & $15.02\pm 0.05$ [mag]\\
        $d_{\odot}$ & $10.07 \pm 0.36$ [kpc] \\
        $R_{G}$ & $8.29\pm 0.20$ [kpc] \\
        $\rm{[M/H]}$ & $-0.4 \pm 0.2$ \\
        $\rm{[Fe/H]}_{spec}$ & $-0.52\pm 0.2$\\ 
        Age & $5\pm 2$ [Gyr]\\
        $\mu_{\alpha}^{\ast}$ & $-7.08 \pm 0.52$ [mas yr$^{-1}$]\\
        $\mu_{\delta}$ & $-1.36 \pm 0.48$ [mas yr$^{-1}$] \\
        $V_{T}^{RA}$ & $-338.04\pm 10.12$  [km s$^{-1}$] \\
        $V_{T}^{Dec}$ & $-64.98\pm 2.14$ [km s$^{-1}$] \\
        RV & $28.3\pm 0.7$ [km s$^{-1}$]  \\
        $r_c$ & $5.5 \pm 0.6$ [pc] \\
        c & 0.73 \\
        $M_{Ks}$ & $-7.33\pm 1.1$ [mag]\\
        $M_{V}$ &  $-5.05$ [mag]\\ 
        e & $0.12\pm 0.7$\\
        $|Z_{max}|$  & $0.6\pm0.3$ [kpc]\\
        $r_{apo}$ & $9.4\pm 1.9$ [kpc]\\
        $r_{peri}$ & $7.6\pm 0.6$ [kpc]\\
\hline\hline
\end{tabular}
\label{table}
\end{table}

\section{Physical parameter determination}
\label{sec:physicalparm}
To determine the main physical parameters, we adopted the same methodology applied in our previous work \citep[e.g.][]{garro22b, garro22c}. %Second, we employed the \texttt{SIESTA} code (Statistical matchIng between rEal and Synthetic sTellar popuLations; \citealt{10.1093/mnras/stae2055}), which is specifically designed to perform statistical isochrone fitting on CMDs of single stellar populations. We explain briefly below the main steps followed.\\

In the NIR, we derived reddening and extinction using two independent approaches. First, we exploited the position of the red clump (RC) standard candles, located at $K_{s} = 13.8$ and $G = 17.8$. Using the empirical photometric calibration of \cite{2018A&A...609A.116R}, we adopted their $K_s$ absolute magnitude and intrinsic color, obtaining a reddening of $E(J-K_s) = 0.44$. This value translates into an extinction of $A_{Ks} = 0.33$, assuming an extinction coefficient of $R_{Ks} = 0.75$ \citep{1989ApJ...345..245C}. From these values, we derived a distance modulus of $(m-M)_0 = 15.08$, corresponding to a heliocentric distance of $d_{\odot} = 10.36\pm 0.52$ kpc.
As a second approach, we used the reddening maps from \cite{2011ApJ...737..103S}, from which we obtained an extinction value of $A_V = 4.15$. Adopting the standard extinction relations, -- $A_{Ks} = 0.11 \times A_V$, $A_G = 0.86 \times A_V$, and $A_G = 2.0 \times E(BP-RP)$ -- we derived $A_{K_s} = 0.46$, $E(J-K_s) = 0.61$, and a distance modulus of $(m-M)_0 = 14.95$, corresponding to a heliocentric distance of $d_{\odot} = 9.77\pm 0.46$ kpc.\\
In the optical, we adopted the relations from \cite{Babusiaux_2018}, obtaining $E(BP-RP) = 1.20$, $A_{G} = 2.40$, and a distance modulus of $(m-M)_0 = 15.02$, corresponding to a heliocentric distance of $d_{\odot} = 10.08\pm 0.81$ kpc.
We then estimated the final heliocentric distance as the mean of the three methods, obtaining $d_{\odot} = 10.07 \pm 0.36$ kpc, where the uncertainty is propagated from the individual estimate. Assuming $R_{\odot} = 8.27$ kpc \citep{GRAVITY2021}, this corresponds to a Galactocentric distance of $R_{G} = 8.29$ kpc and a vertical height above the Galactic plane of $Z = 1.59$ kpc.\\

Using the mean cluster PMs and distance, we computed the tangential velocity components, obtaining $V_{T}^{RA} \approx -338.048\pm 10.12$ km s$^{-1}$ and $V_{T}^{Dec} \approx -64.98\pm 2.14$ km s$^{-1}$. These positional and kinematic properties collectively indicate that Garro~04 is a disk cluster.\\

\begin{figure*}[htpb]
\sidecaption
\begin{minipage}{12cm}
\includegraphics[width=12cm]{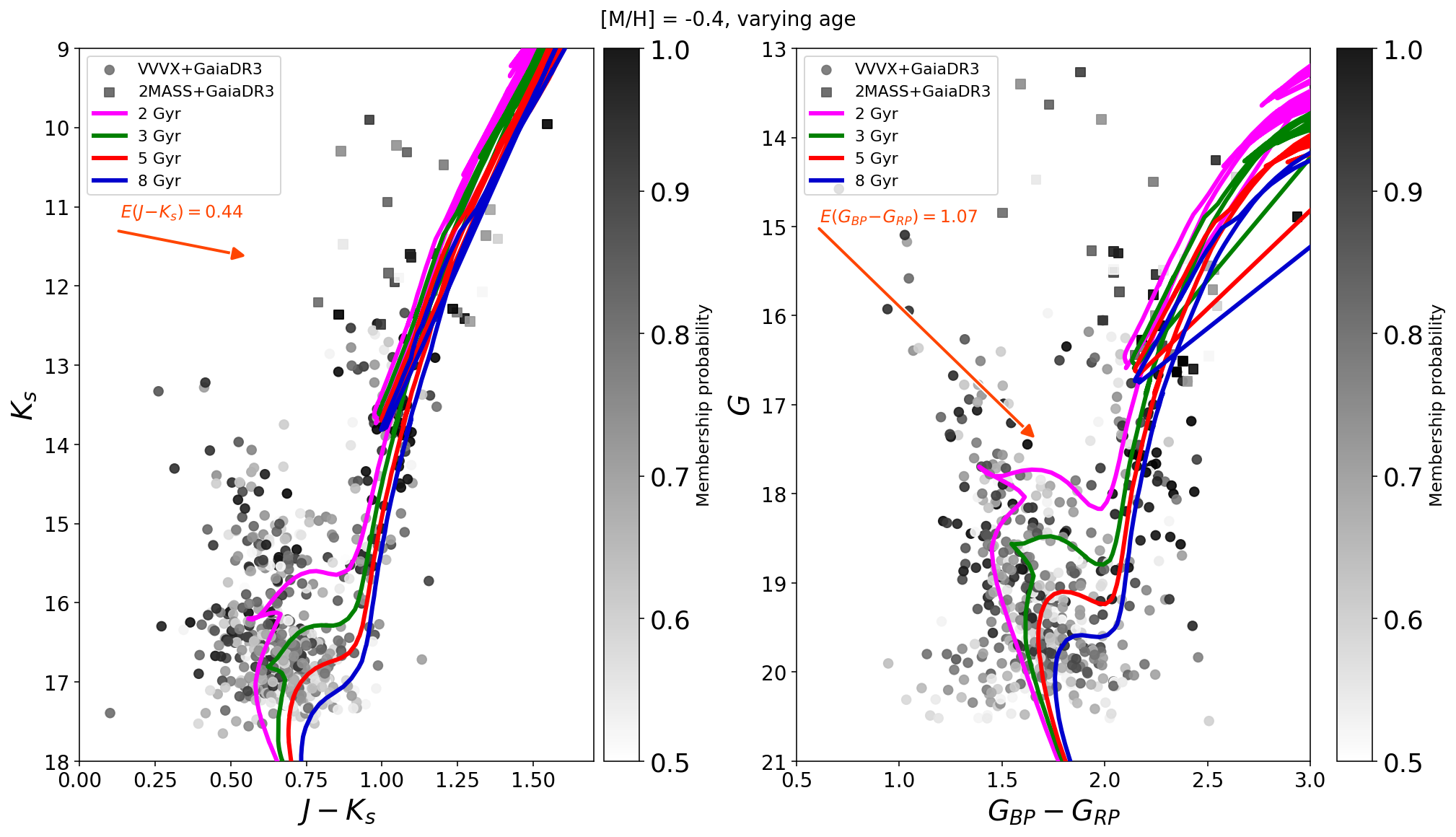}\\[2pt]
\includegraphics[width=12cm]{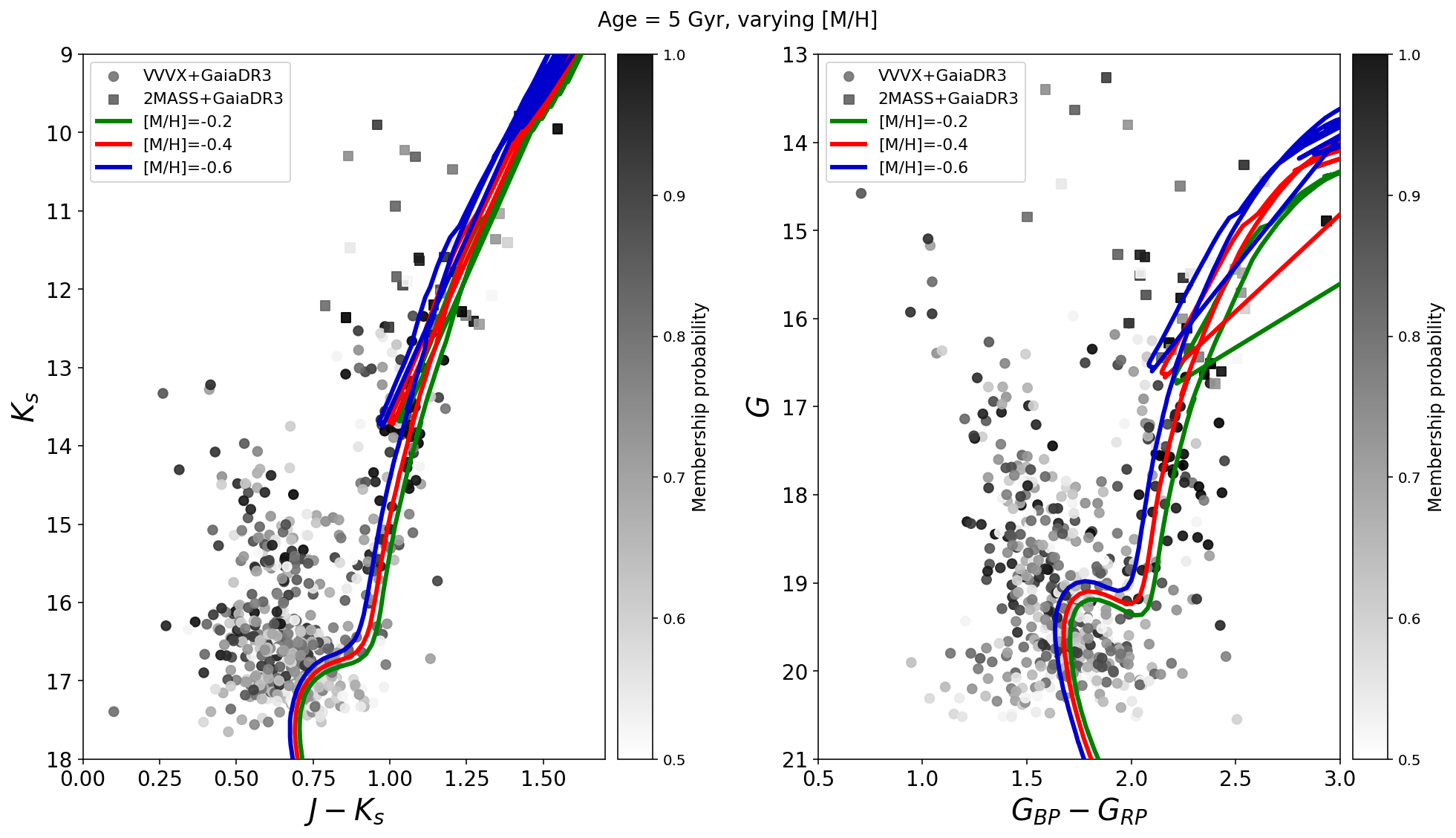}
\end{minipage}
\caption{Near-infrared (\textit{left}) and optical (\textit{right}) CMDs of Garro~04. Circle and square symbols represent stars from the VVVX+{\it Gaia}~DR3 
and 2MASS+{\it Gaia}~DR3 catalogs, respectively, color-coded by membership probability ($p > 0.5$). PARSEC isochrones are overplotted 
for a fixed distance modulus and reddening, exploring a range of ages (3, 5, and 8~Gyr) at fixed metallicity ($\rm{[M/H]} = -0.4$) in the upper panels, and a range of global metallicities ($\rm{[M/H]} = -0.2, -0.4, -0.6$) at fixed age (5~Gyr) in the bottom panels. The orange arrows draw the reddening vector in both optical and near-infrared CMDs.}\label{fig:cmd_varyingagemet}
\end{figure*}

Once reddening, extinction, and the distance modulus were established, we used these parameters to constrain the metallicity and age through isochrone fitting. A major limitation is the well-known age–metallicity degeneracy, which becomes particularly severe when the main-sequence turn-off (MSTO) is poorly sampled. In our case, only the very upper portion of the MSTO is accessible in the VVVX+\textit{Gaia} data, significantly reducing the robustness of the age determination.
As shown in Fig. ~\ref{fig:cmd_varyingagemet}, we explored a grid of solutions in both the NIR and optical bands, using PARSEC+COLIBRI isochrones\footnote{\url{https://stev.oapd.inaf.it/cgi-bin/cmd}; input form version 3.8.}
\citep[][version 1.2S]{Bressan2012parsec} with solar-scaled models and no rotation. The best-fitting PARSEC models yield an age of $5 \pm 2$ Gyr and a metallicity of $\rm{[M/H]}= -0.4\pm 0.2$. Adopting the following equation \ref{equation} \citep{1993ApJ...414..580S,2004ApJ...616..498C}

\begin{equation}
\label{equation}
    \rm{[M/H] = [Fe/H] + \log(0.694 \times 10^{[\alpha/Fe]} + 0.301)}.
\end{equation} 
we converted $\rm{[M/H]}$ into iron abundance assuming $\rm{[\alpha/Fe]}=+0.20$, yielding $\rm{[Fe/H]}=-0.55\pm0.20$.

\section{ESO/KMOS observations and radial velocity determination}
\label{sec:kmosdata}

Garro~04 was observed as part of the ESO Public Spectroscopic Survey VVVX-GalCen (Run ID 116.29EF.001; PI: Matias G\'omez) with the K-band Multi Object Spectrograph (KMOS), a near-infrared spectrograph mounted on VLT UT1 (Antu) at the ESO Paranal Observatory.

\begin{figure*}[htpb]
\centering
\includegraphics[width=\textwidth]{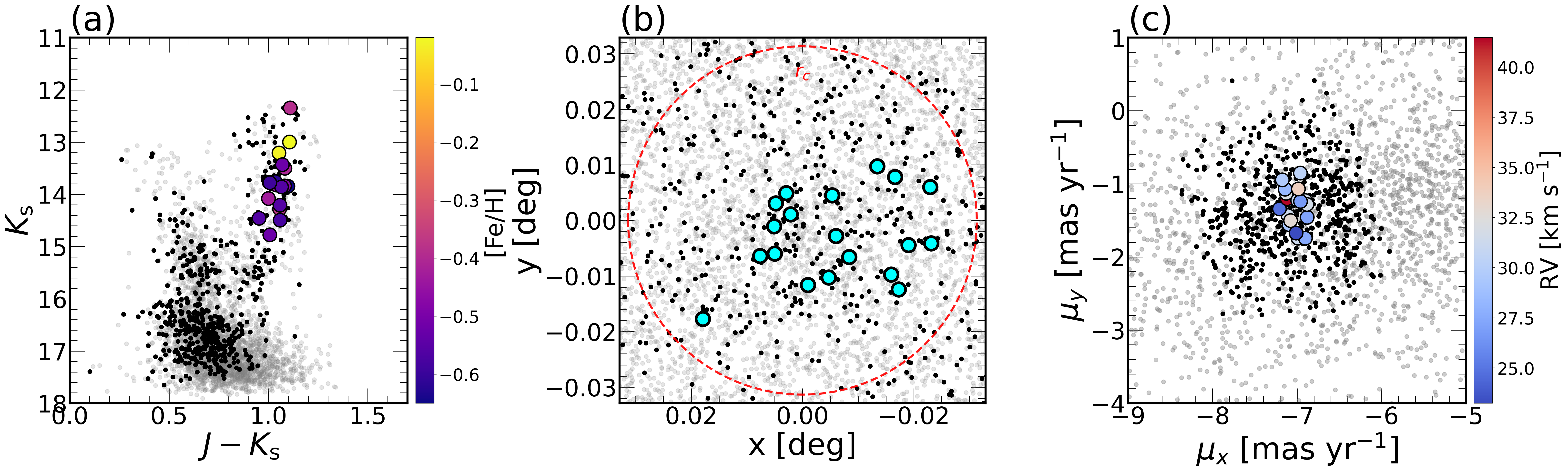}
\caption{KMOS target selection. (a) VVVX colour–magnitude diagram (CMD) for field stars selected within $\sim3'$ of the cluster center (grey points) and high-probability proper-motion (PM) cluster members (black points). KMOS targets are colour-coded according to their spectroscopic metallicity, $[\mathrm{Fe/H}]_{\rm spec}$. (b) Spatial distribution of the two samples in the $(x,y)$ coordinate system, following Equation (2) of \cite{2018A&A...616A..12G}. The red circle indicates the core radius, $r_{\rm c}\sim1.88'$ and cyan points are the KMOS targets. (c) Vector proper-motion diagram in the $(x,y)$ coordinate system. KMOS targets are colour-coded according to their radial velocities (RVs).}
\label{fig:3panels}
\end{figure*}

We selected 19 candidate member stars, while 3 integral-field-units (IFUs) were positioned on blank sky regions for sky subtraction. By design, only red giant stars with $K_s \approx 12-15$ and highest PM membership probability (>80\%) were selected. Also, the target selection was further optimized to maximize the allocation efficiency of the KMOS IFUs. Owing to the 7.5 arcmin field of view of KMOS, all stars with measured radial velocities are located well within the cluster core, as displayed in Figure \ref{fig:3panels}. The observations were obtained with a single KMOS pointing using the \textit{H}-band grating configuration. A total exposure time of 60 min was acquired, beginning at UT 01:43:14 on 11${\rm th}$ March 2026 under seeing conditions of approximately $<$0.98 arcsec.

Spectra were reduced and calibrated using the ESO Data Processing System \citep[EDPS;][]{Freudling2024} and the KMOS pipeline version 4.5.3. 
Radial velocities for Garro~04 were measured using KMOS Radial Velocities via Cross-correlation with the VVV Survey (KROSS), a custom Python pipeline developed for VLT/KMOS data from the VVV-GalCen Survey. KROSS derives heliocentric radial velocities by cross-correlating continuum-normalised $H$-band spectra with PHOENIX synthetic templates degraded to the KMOS resolution ($R\sim4000$; \citealt{2013A&A...553A...6H}). The pipeline adopts the velocity corresponding to the peak of the cross-correlation function after applying barycentric corrections. This methodology was validated using observations of the Galactic GC M22; the results are in excellent agreement with independent literature measurements. A detailed description of the pipeline and its validation will be presented in Obasi et al. (in prep.).
The signal-to-noise ratio ($S/N$) of each spectrum was estimated empirically as the ratio between the median flux level and the standard deviation of the residual spectrum after subtracting a median-filtered continuum model. The final $S/N$ ratios ranging from 16 to 28 pixel$^{-1}$. Figure \ref{figOrbits} shows the $S/N$-weighted stacked spectrum of the Garro~04 member stars together with the four highest-$S/N$ individual spectra, compared to their best-fitting PHOENIX templates shifted to the adopted systemic velocity. The excellent agreement between the observed and synthetic spectra across the $H$-band, particularly around the prominent Mg\,I, Si\,I, and Fe\,I absorption features, demonstrates that the radial-velocity solution is constrained by multiple stellar lines rather than individual spectral features or local noise fluctuations. \\

Given the moderate spectral resolution of KMOS ($R\approx4,000$), the data are not suitable to obtain precise elemental abundances. However, they allow radial velocities to be measured with an accuracy of a few km~s$^{-1}$, sufficient to assess membership and characterize the kinematics of Garro~04.

All observed targets are associated with a well-defined radial-velocity peak and are therefore considered bona fide members of Garro~04 (see Table~\ref{table:RV}). From this sample, we derive a weighted mean cluster radial velocity of $29.24 \pm 0.25$ km\,s$^{-1}$. We note that three stars deviate from the cluster mean, but by no more than $1.5\sigma$; they are therefore still consistent with membership and are retained as members.

The member stars exhibit a velocity dispersion of $\sigma = 3.60 \pm 0.88$ km s$^{-1}$ (without the outliers with RV $=41.5$ km s$^{-1}$, $\sigma = 2.69 \pm 0.48$ km s$^{-1}$), which is somewhat lower than the values typically measured for star clusters observed as part of the VVVX-GalCen survey. However, given the relatively small sample size and the moderate spectral resolution of the data, this estimate should be interpreted with caution until additional radial-velocity measurements become available. 

\begin{table*}[htpb]
%\onecolumn
\centering 
\renewcommand{\arraystretch}{1.0}
\caption{KMOS targets observed in the field of Garro~04. The table lists the {\it Gaia} source identifier, equatorial coordinates, radial velocity, signal-to-noise ratio ($S/N$), and membership classification for each observed target. Asterisks mark the three stars with radial velocities lower than the peak of the cluster distribution. These stars remain consistent with membership, as their velocities are within the estimated uncertainties within $1.5\sigma$ of the cluster systemic velocity.}
\begin{tabular}{lccccccccl}
\hline\hline
        {\it Gaia} ID          &   RA  & DEC       & $G$ & $K_{s}$ & $\mu_{\alpha}$ & $\mu_{\delta}$ & RV & S/N & Membership \\
                               & [deg] & [deg] & [mag] & [mag] & [mas yr$^{-1}$] & [mas yr$^{-1}$] & [km s$^{-1}$ ] & & \\
\hline
             
  5864083436772875264 & 203.089 & -64.671 &  16.99 &   12.99 &  -6.921 & -1.134     & $30.34 \pm 1.41$ & 21.05 & Member    \\ 
  5864083810333084544 & 203.065 & -64.660 &  16.51 &   12.35 &  -7.108 & -1.252     & $32.00 \pm 1.14 $ & 26.65 & Member    \\ 
  5864083810355782784 & 203.059 & -64.659 &  17.58 &   13.77 &  -7.128 & -1.154     & $28.35 \pm 1.33$ & 24.00& Member    \\ 
  5864083814699216000 & 203.045 & -64.665 &  17.87 &   13.86 &  -6.996 & -1.211     & $30.91 \pm 2.14$ & 22.22  & Member    \\ 
  5864083844715395968 & 203.007 & -64.666 &  17.09 &   13.21 &  -6.965 & -1.472     & $27.49 \pm 0.84$ & 22.35 & Member    \\ 
  5864083844715404928 & 203.010 & -64.663 &  17.90 &   13.84 &  -6.943 & -0.978     & $31.77 \pm 0.98$ & 23.41 & Member    \\ 
  5864083879053770240 & 203.036 & -64.664 &  18.15 &   14.21 &  -6.896 & -1.209     & $31.59 \pm 1.42$ & 19.27 & Member    \\ 
  5864083879053774080 & 203.028 & -64.660 &  18.47 &   14.50 &  -7.138 & -1.067     & $41.47 \pm 1.48$ & 20.43 & Member    \\ 
  5864083879053781504 & 203.059 & -64.655 &  17.55 &   13.78 &  -6.944 & -1.320     & $26.67  \pm 0.81$ & 28.29 & Member (*)\\ 
  5864083879062387328 & 203.052 & -64.652 &  18.52 &   14.78 &  -6.979 & -1.103     & $33.80\pm 1.67 $ & 17.57 & Member    \\ 
  5864083879082757760 & 203.054 & -64.649 &  17.71 &   14.08 &  -7.001 & -1.719     & $23.24 \pm 0.73$ & 22.00 & Member (*)\\ 
  5864083883418740352 & 203.033 & -64.656 &  18.23 &   14.46 &  -6.899 & -1.364     & $27.21 \pm 1.48$ & 16.13& Member    \\ 
  5864084050852466688 & 202.993 & -64.658 &  17.98 &   13.85 &  -7.070 & -1.375     & $31.22  \pm .84$ & 23.84& Member    \\ 
  5864084055217451136 & 203.008 & -64.646 &  18.20 &   14.28 &  -7.254 & -1.070     & $24.80 \pm 0.85$ & 20.10 & Member (*)\\ 
  5864084055248141440 & 203.003 & -64.658 &  17.66 &   13.76 &  -7.135 & -1.196     & $32.84  \pm 0.82 $ & 21.48 & Member    \\ 
  5864084085212562176 & 203.016 & -64.644 &  17.76 &   13.82 &  -7.147 & -1.217     & $31.37  \pm 1.26$ & 22.76 & Member    \\ 
  5864084089577238272 & 203.035 & -64.649 &  17.69 &   13.80 &  -7.113 & -1.471     & $29.23  \pm 1.07$ & 20.73& Member    \\ 
  5864084123967628928 & 202.994 & -64.648 &  17.59 &   13.50 &  -6.999 & -1.388     & $32.83 \pm 1.61$ & 22.36  & Member    \\ 
  5864084261375980928 & 203.058 & -64.650 &  17.36 &   13.43 &  -7.163 & -1.023     & $29.70 \pm 1.26 $ & 27.82 & Member    \\ 
\hline\hline
\end{tabular}
\label{table:RV}
\end{table*}

\section{Orbital elements of Garro~04}
\label{sec:orbits}

\begin{figure*}[htpb]
\centering
\includegraphics[width=0.9\textwidth]{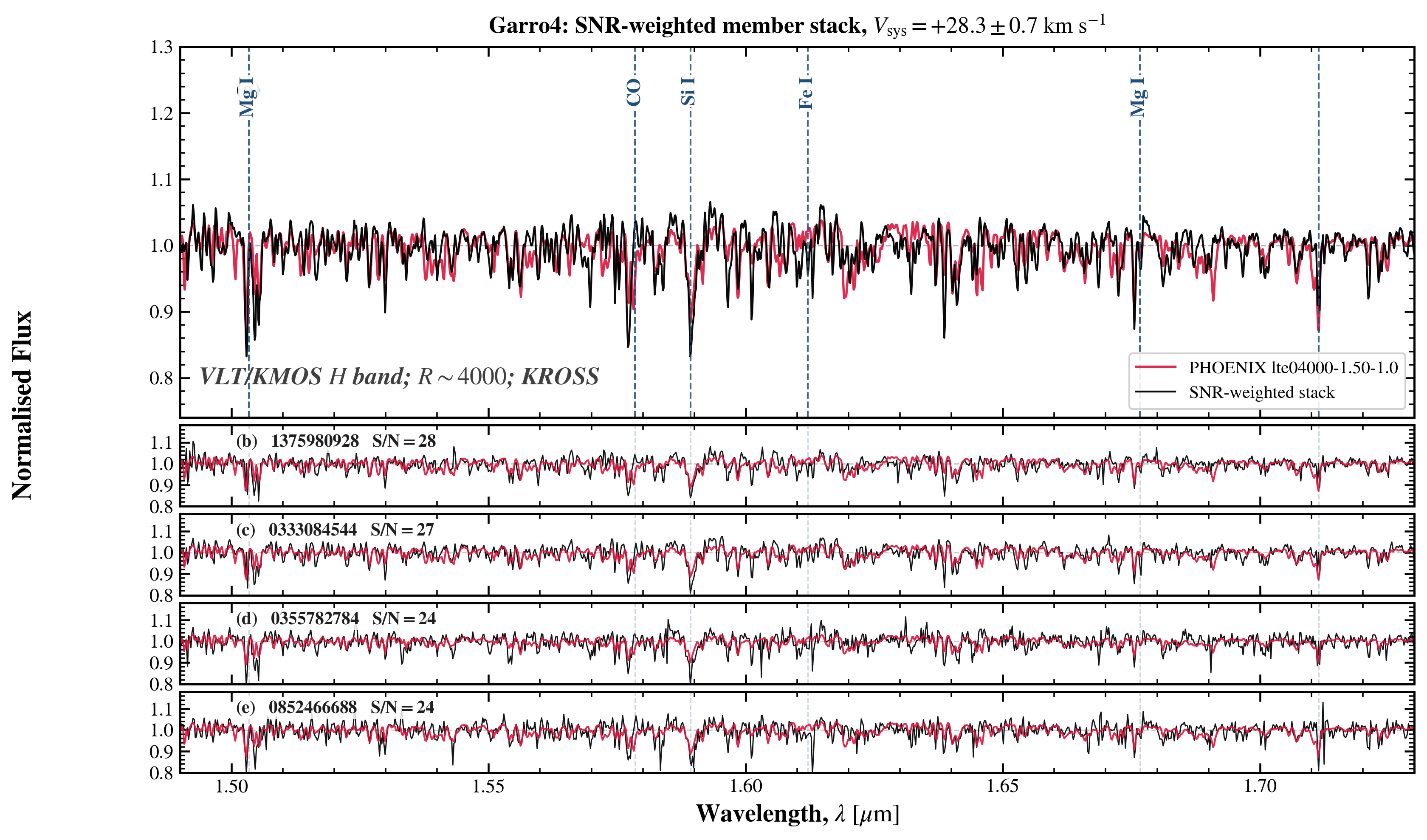}
\includegraphics[width=0.9\textwidth]{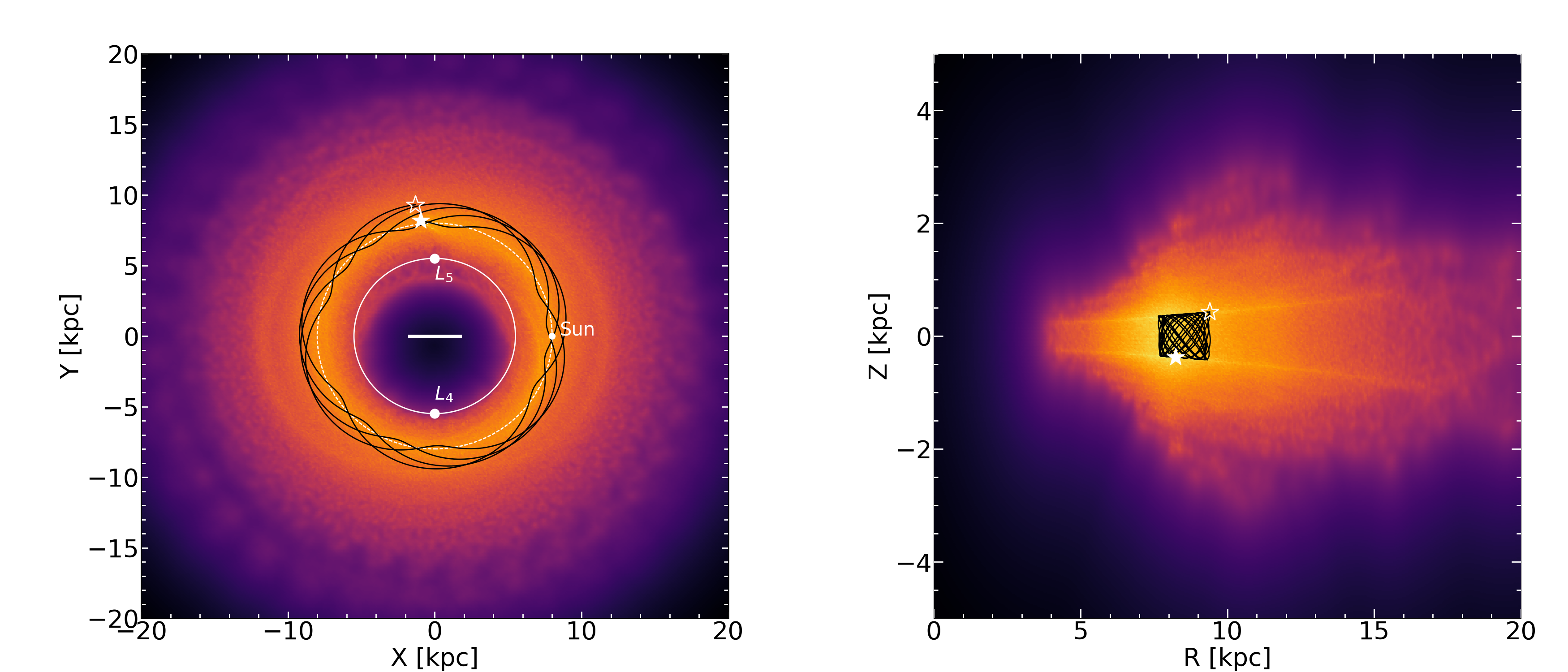}
\caption{\textbf{Upper panels.} Representative KMOS member spectra and PHOENIX template comparison for Garro~04. Panel \textit{(a)} shows the $S/N$-weighted stacked spectrum of all adopted cluster members (black) together with the best-matching PHOENIX synthetic template (red), shifted to the adopted cluster systemic velocity ($V_{\rm sys}=+28.3\pm0.7$ km s$^{-1}$). Panels \textit{(b)–(e)} show the four highest-$S/N$ member stars with their corresponding best-fitting PHOENIX templates. Vertical dashed lines indicate the expected positions of prominent absorption features, including Mg I, CO, Si I, and Fe I transitions used in the RV analysis. The consistency of these features demonstrates the reliability of the member selection and systemic RV. \textbf{Bottom panels.} Ensemble of one million orbits in the frame co-rotating with the bar for the Garro~04 cluster, projected on the equatorial (\textit{left}) and meridional (\textit{right}) Galactic planes in the non-inertial reference frame with a bar pattern speed of 41~km~s$^{-1}$~kpc$^{-1}$, and time-integrated backward over 2 Gyr. The yellow and orange colors correspond to more probable regions of the space, which are most frequently crossed by the simulated orbits. The solid inner white and outer dashed circles in the \textit{left} panel show the locations of the CR (see text) and the solar orbit, respectively. The while dots mark the positions of the Lagrange points of the Galactic bar, $L_{4}$ and $L_{5}$, and the current position of the Sun, respectively. The horizontal solid white line shows the extension of the bar \citep[R$_{\rm c}\sim 3.28$ kpc;][]{Robin2012} in our model. The black line shows the orbit of Garro~04, from the observables without error bars. The white-filled and unfilled star symbols indicate the initial and final positions of Garro~04 in our simulations, respectively.}
\label{figOrbits}
\end{figure*}

We made use of the state-of-art Milky Way model \texttt{GravPot16}\footnote{\url{https://gravpot.utinam.cnrs.fr}} to predict the orbital path of Garro~04 in a steady-state gravitational Galactic model that includes a ``boxy/peanut" bar structure \citep{Fernandez-Trincado2017, Fernandez-Trincado2019}. Orbits have been integrated with the \texttt{GravPot16}, which includes the perturbations due to a realistic (as far as possible) rotating ``boxy/peanut" bar, which fits the structural and dynamical parameters of the Galaxy to the best we know of the recent knowledge of our Milky Way. 

For the orbit computations, we adopt the same model configuration, solar position and velocity vector as described in \citet{Fernandez-Trincado2019}, except for the angular velocity of the bar ($\Omega_{\rm bar}$), for which we employed the recommended value of 41 km s$^{-1}$ kpc$^{-1}$ \citep[][]{Sanders2019}, and by assuming variations of $\pm$10 km s$^{-1}$ kpc$^{-1}$. 

The considered structural parameters of our bar model (e.g., mass and orientation) are within observational estimations that lie in the range of 1.1$\times$10$^{10}$ M$_{\odot}$ and present-day orientation of 20$^{\circ}$ \citep{Fernandez-Trincado2017} in the non-inertial frame (where the bar is at rest). The bar scale lengths are $x_0=$1.46 kpc, $y_{0}=$ 0.49 kpc, and $z_0=$0.39 kpc, and the middle region ends at the effective semi-major axis of the bar $Rc = 3.28$ kpc \citep{Robin2012}. Our bar model locates the corotation radius (CR) at 5.5 kpc for $\Omega_{\rm bar}= 41$ km s$^{-1}$ kpc$^{-1}$.

For guidance, our Galactic convention is as follows: $X-$axis is oriented toward $l=$ 0$^{\circ}$ and $b=$ 0$^{\circ}$, $Y-$axis is oriented toward $l$ = 90$^{\circ}$ and $b=$0$^{\circ}$, and the disk rotates toward $l=$ 90$^{\circ}$; the velocity is also oriented in these directions. Following this convention, the Sun's orbital velocity vectors are [U$_{\odot}$, V$_{\odot}$, W$_{\odot}$] = [$11.1$, $12.24$, 7.25] km s$^{-1}$ \citep{Ralph2010}. The model has been rescaled to the Sun's Galactocentric distance, 8.27 kpc \citep{GRAVITY2021}, and the circular velocity at the solar position to be $\sim$ $229$ km s$^{-1}$ \citep{Eilers2019}.

The most likely orbital parameters and their uncertainties are estimated using a simple Monte Carlo scheme. An ensemble of one million orbits was computed backward in time for 2 Gyrs, under variations of the observational parameters assuming a normal distribution for the uncertainties of the input parameters (e.g., positions, heliocentric distances, radial velocities, and PMs), which were propagated as 1$\sigma$ variations in a Gaussian Monte Carlo resampling. To compute the orbits of Garro~04, we adopt a mean RV (Table \ref{table}), including only member stars listed in Table \ref{table:RV}. The nominal PMs have been taken from Table \ref{table}, with an assumed uncertainty of 0.3 mas yr$^{-1}$ for the orbit computations. The heliocentric distance ($d_{\odot}$) was adopted from Table \ref{table}. The results for the main orbital elements are listed and discussed below.

Figures \ref{figOrbits} show the simulated orbits adopting a simple Monte Carlo approach for Garro~04. The probability densities of the resulting orbits projected on the equatorial and meridional galactic planes, in the non-inertial reference frame where the bar is at rest are highlighted at the same figure. The black line in the figure shows the orbital path for Garro~04 (adopting observables without uncertainties). The yellow color corresponds to the most probable regions of the space, which are crossed more frequently by the simulated orbits. For each orbit in Fig. \ref{figOrbits}, we calculate the perigalactocentric distance ($r_{\rm peri}$), the apogalactocentric distance ($r_{\rm apo}$), the maximum vertical excursion from the Galactic plane (|Z$_{\rm max}$|) and the orbital eccentricity defined as ($r_{\rm apo}$ $-$ $r_{\rm peri}$)/($r_{\rm apo}$ $+$ $r_{\rm peri}$). The median value of the orbital elements was found for one million realizations, with uncertainty ranges given by the 16$^{\rm th}$ and 84$^{\rm th}$ percentile values. Following the same strategy as described by \cite{Fernandez-Trincado2022}, we calculated the z-component of the angular momentum in the inertial frame to know whether the orbital motion of Garro~04 has a prograde and/or retrograde sense with respect to the rotation of the bar. Since this quantity is not conserved in a model with bar- and/or spiral-arm structures, we were interested only in the sign. 

Figure \ref{figOrbits} clearly shows that Garro~04 is confined in a disk-like orbital configuration that lies in an in-plane orbit with low eccentricity, $\lesssim0.12\pm0.07$, and low vertical (|Z$_{\rm max}|\lesssim0.6\pm0.3$ kpc) from the Galactic plane, and $r_{\rm apo}\sim9.4\pm1.9$ kpc slightly larger than the Solar orbit, and the $r_{\rm peri}\sim7.6\pm0.6$ kpc slightly lower than the Solar orbit, and it is confined to a prograde orbit. Orbits in Garro~04 have energies that allow the cluster to move trapped around the Solar orbit. We find that Garro~04 is a cluster confined within the solar orbit that lives in the disk due to possible effects of orbital trapping due to resonances created by the bar on the Galactic plane \citep{Moreno2021}.

\section{KMOS metallicity determination}
\label{sec:specmet}
We used KROSS code to estimate metallicities on a star-by-star basis for individual cluster members. Here, we briefly describe the procedure used to construct and validate the metallicity model; a more detailed description of the methodology will be presented in Obasi et al. (in prep.). KROSS was trained on a sample of field red giant branch (RGB) stars from APOGEE. We first constructed an empirical training set from continuum-normalized APOGEE DR17 spectra and their corresponding ASPCAP $\rm{[Fe/H]}$ measurements \citep{2016AJ....151..144G,2022ApJS..259...35A}. We selected stars with valid ASPCAP parameters, no \texttt{STAR\_BAD} flag, and APOGEE $S/N \geq 120$. Because our KMOS GC targets are predominantly cool giants, we restricted the training sample to $3800 \leq T_{\rm eff} \leq 5000$ (K), $0.5 \leq \log g \leq 3.2$ (cgs), and $-2.10 \leq \rm{[Fe/H]} \leq +0.55$ (dex). These cuts produced an initial sample of 8172 stars.\\
The APOGEE catalogue is not uniformly distributed across this parameter space, so a simple random selection would strongly favour the most densely populated regions. To obtain broader coverage, we sampled the stars jointly in [Fe/H], $T_{\rm eff}$, and $\log g$ using cells of 0.25 dex in metallicity, 300 K in effective temperature, and 0.75 dex in surface gravity, selecting up to three stars from each occupied cell. This produced a final full-spectrum training sample of 294 stars: 177 observed with the Apache Point Observatory (APO) 2.5-m telescope and 117 with the Las Campanas Observatory (LCO) 2.5-m telescope. The parameter coverage and metallicity distribution of the final training sample are summarized in Table \ref{table:apogee}. Of the 294 stars, 117 have [Fe/H]~$> -0.5$. We used this subsample to examine the model performance separately in the relatively metal-rich regime, which is particularly relevant for many of the inner-Galaxy KMOS targets.

\begin{table}[htpb]
\centering
\renewcommand{\arraystretch}{1.1}
\caption{Training sample. Parameter coverage and metallicity distribution of the APOGEE training sample (N = 294).}
\begin{tabular}{lccc}
\hline\hline
Parameter & Minimum & Median & Maximum \\
\hline
$T_{\rm eff}$ [K]        & 3800     & 4399     & 4949     \\
$\log g$ [dex]           & 0.52     & 1.74     & 3.19     \\
$\mathrm{[Fe/H]}$ [dex]  & $-2.099$ & $-0.751$ & $+0.546$ \\
APOGEE $S/N$             & 120      & 198      & 1055     \\
\hline
\multicolumn{4}{c}{Metallicity distribution} \\
\hline
\multicolumn{2}{l}{APOGEE $\mathrm{[Fe/H]}$ range [dex]} & \multicolumn{2}{c}{Number of stars} \\
\hline
\multicolumn{2}{l}{$-2.10$ to $-2.00$}          & \multicolumn{2}{c}{18}  \\
\multicolumn{2}{l}{$-2.00$ to $-1.50$}          & \multicolumn{2}{c}{46}  \\
\multicolumn{2}{l}{$-1.50$ to $-1.00$}          & \multicolumn{2}{c}{46}  \\
\multicolumn{2}{l}{$-1.00$ to $-0.50$}          & \multicolumn{2}{c}{67}  \\
\multicolumn{2}{l}{$-0.50$ to $\phantom{+}0.00$} & \multicolumn{2}{c}{58}  \\
\multicolumn{2}{l}{$\phantom{+}0.00$ to $+0.50$} & \multicolumn{2}{c}{47}  \\
\multicolumn{2}{l}{$+0.50$ to $+0.55$}          & \multicolumn{2}{c}{12}  \\
\hline
\multicolumn{2}{l}{Total}                       & \multicolumn{2}{c}{294} \\
\hline\hline
\end{tabular}
\label{table:apogee}
\end{table}

\subsection{Spectral preprocessing}
The original APOGEE spectra have a resolving power of $R=22500$ \citep{2019PASP..131e5001W}. To make the training spectra more representative of the KMOS observations, we convolved them with a Gaussian kernel to reproduce the approximate KMOS $H-$band resolving power of 4000. This corresponds to an additional Gaussian broadening with a FWHM of approximately 73.8 km s$^{-1}$. The degraded spectra were then interpolated onto a common KMOS-like wavelength grid covering $1.5115 \leq \lambda \leq 1.6990$ $\mu m$ with a pixel spacing of 2.17 $\AA$. Before convolution, pixels affected by APOGEE bad-pixel, persistence, skyline, or related quality flags were masked. Each spectrum was then continuum-normalized using a fourth-order Legendre polynomial with iterative asymmetric sigma clipping. We applied the same normalization procedure to the KMOS target spectra to keep the treatment of the training and science spectra as consistent as possible.

\subsection{Noise augmentation}
Although all selected APOGEE spectra have $S/N \geq 120$, the KMOS observations generally have considerably lower $S/N$. We therefore added artificial noise to the training spectra so that the model was exposed to spectra more representative of the KMOS data. For each APOGEE spectrum, we generated noise realizations corresponding to $S/N = 15,\ 25,\ 40$, and 80. In the cross-validation analysis, the held-out validation spectra were evaluated at $S/N = 25$. All augmented versions of a given star were kept within the training portion of the same fold, ensuring that no noisy realization of a validation star appeared in the corresponding training set.

\subsection{Full spectrum metallicity model}
We used the information contained across the full $H-$band spectrum. The normalized spectra were first standardized pixel by pixel and then compressed using principal-component analysis (PCA). A ridge-regression model was subsequently trained to predict the APOGEE DR17 ASPCAP [Fe/H] values from the PCA coefficients. We tested $N\_PCA = 3, 5, 8$, and 12 principal components together with ridge-regularization parameters between $10^{-5}$ and 10. The selected spectrum-only model used 12 principal components and $\alpha= 10^{-5}$. To avoid information leakage, all preprocessing steps, including missing-pixel replacement, pixel scaling, PCA decomposition, and regression fitting were performed independently within each training fold. We used five-fold cross-validation, stratified in 0.25-dex metallicity intervals, to maintain broad [Fe/H] coverage across the individual folds.

\subsection{Validation results and its application to the Garro 04 stars}
\begin{figure*}[htpb]
\centering
\includegraphics[width=0.9\textwidth]{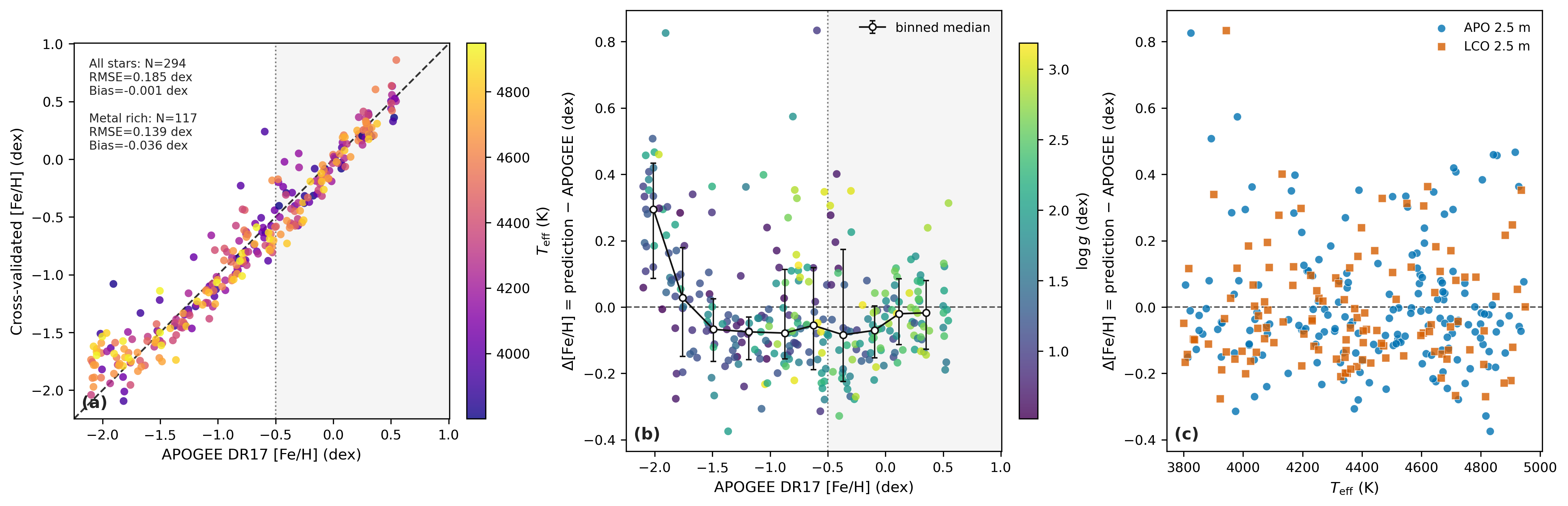}
\caption{Held-out cross-validation performance of the augmented full-spectrum model, in which APOGEE $T_{\rm eff}$ and $\log g$ are included alongside the spectral principal-component coefficients. (a) Cross-validated predicted metallicities compared with the APOGEE DR17 reference values; the dashed line marks the one-to-one relation. (b) Metallicity residuals, $\Delta \rm{[Fe/H]} = \rm{[Fe/H]}_{\rm prediction} -\rm{[Fe/H]}_{\rm APOGEE}$, as a function of APOGEE metallicity; the binned median is shown for reference. (c) Metallicity residuals as a function of $T_{\rm eff}$, with the APO 2.5-m and LCO 2.5-m subsamples shown separately. The augmented model is shown as a performance comparison; the final KMOS metallicities were obtained with the spectrum-only model.}
\label{fig:validation}
\end{figure*}

The spectrum-only model was designed to predict metallicity directly from the normalized spectrum and therefore does not require independently determined $T_{\rm eff}$ or $\log g$ at the application stage. Across the complete 294-star validation sample, this model achieved an RMSE of 0.193 dex, a mean residual of -0.002 dex, and a robust residual scatter of 0.162 dex. Its performance improved for the 117 metal-rich stars with $\rm{[Fe/H]} > -0.5$, for which the RMSE was 0.153 dex and the mean residual was -0.043 dex. For comparison, we trained an augmented version of the model in which the APOGEE $T_{\rm eff}$ and $\log g$ values were included as additional predictors alongside the spectral principal-component coefficients. The cross-validation performance of this augmented model is shown in Figure \ref{fig:validation}. Figure \ref{fig:validation}(a) compares the cross-validated metallicity predictions with the APOGEE DR17 reference metallicities. The predictions follow the one-to-one relation closely over most of the metallicity range, with an overall RMSE of 0.185 dex and a mean residual of -0.001 dex. For the metal-rich subset, the RMSE decreases to 0.139 dex, with a mean residual of -0.036 dex. Thus, including the atmospheric parameters provides a modest improvement over the spectrum-only model. The residuals as a function of APOGEE metallicity are shown in Figure \ref{fig:validation}(b). A metallicity-dependent residual structure is apparent despite the nearly zero global mean bias. Near the metal-poor boundary, particularly at $\rm{[Fe/H]} \lesssim -1.7$, the predictions are systematically displaced toward higher metallicity. This behaviour is consistent with regression toward the more densely sampled centre of the training distribution. At higher metallicities, the binned median residual remains closer to zero, although substantial star-to-star scatter is still present. Figure \ref{fig:validation}(c) shows the residuals as a function of effective temperature, with the APO 2.5-m and LCO 2.5-m subsamples identified separately. No strong global residual trend with $T_{\rm eff}$ is apparent across the temperature range of the training sample. In addition, the two telescope subsamples broadly overlap, with no clear systematic offset between the APO and LCO observations.
Despite the modest improvement obtained by including $T_{\rm eff}$ and $\log g$, we adopted the APOGEE-trained spectrum-only model for the final KMOS analysis. The atmospheric parameters available for the KMOS stars are inferred from a discrete PHOENIX template grid and may not lie on the same scale as the ASPCAP parameters used during training. Including them as predictors could therefore transfer template-selection errors or parameter-scale differences directly into the metallicity estimates. The adopted spectrum-only model retains the APOGEE-based metallicity calibration while providing a direct mapping from the normalized KMOS spectrum to [Fe/H], without requiring separately inferred atmospheric parameters. As an external validation of the transfer from degraded APOGEE spectra to native KMOS observations, we applied the adopted model to independently selected M22 members, recovering $\rm{[Fe/H]}=-1.67\pm 0.02\,(\rm stat)\pm 0.19\,(\rm cal)$ dex, consistent with the literature metallicity of approximately $\rm{[Fe/H]}\simeq -1.70$ dex (\citealt{1996AJ....112.1487H}, 2010 edition).\\

\begin{figure}[htpb]
\centering
\includegraphics[width=0.5\textwidth]{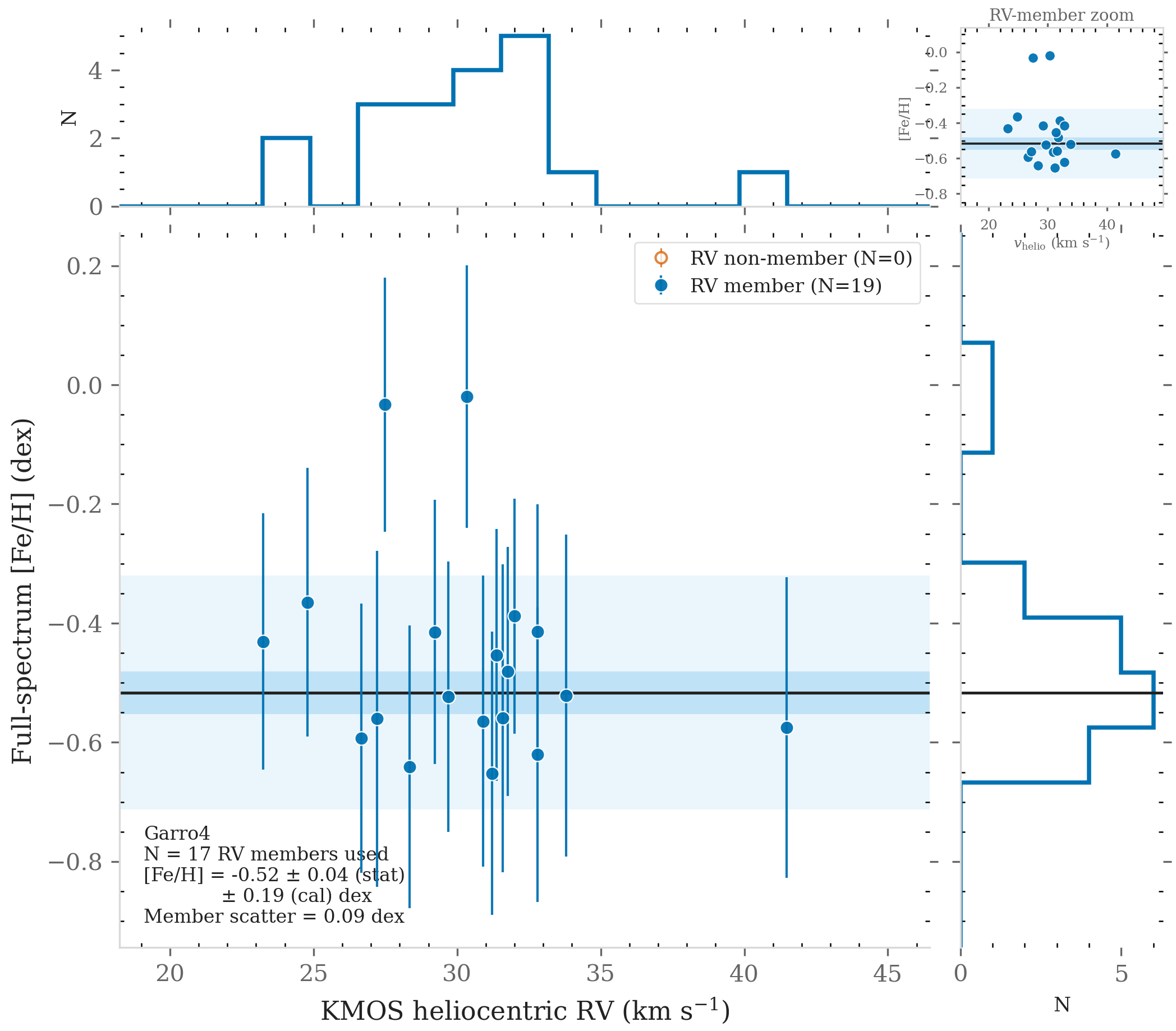}
\caption{Spectroscopic metallicity versus radial velocity for the KMOS targets in Garro~04.
\textit{Main panel:} full-spectrum $\mathrm{[Fe/H]}$ as a function of the KMOS heliocentric radial velocity for each observed star, with the corresponding uncertainties shown as error bars. Filled blue circles mark the RV-selected members ($N=19$); no RV non-members are found. The solid black line indicates the mean cluster metallicity derived from the $N=17$ members retained for the metallicity estimate, $\mathrm{[Fe/H]} = -0.52 \pm 0.04\,(\mathrm{stat}) \pm 0.19\,(\mathrm{cal})$~dex, while the darker and lighter shaded bands show the statistical and calibration uncertainties, respectively. The measured star-to-star scatter of the members is $0.09$~dex.
\textit{Top and right panels:} marginal distributions of the heliocentric radial velocities and of the $\mathrm{[Fe/H]}$ values, respectively.
\textit{Top-right inset:} zoom on the RV members in the $\mathrm{[Fe/H]}$--$v_{\rm helio}$ plane, highlighting the two metal-rich outliers that lie well above the cluster mean.}
\label{fig:Garro04_RV_FeH}
\end{figure}

Hence, by applying this methodology to the Garro~04 stars observed with KMOS, we obtained the individual [Fe/H] estimates listed in Table~\ref{table:metallicity} and shown in Fig.~\ref{fig:Garro04_RV_FeH}. These yield a mean cluster metallicity of $\rm{[Fe/H]} = -0.52 \pm 0.04\,(\mathrm{stat}) \pm 0.19\,(\mathrm{cal})$, in excellent agreement with the photometric estimate. We note, however, that two stars are $\sim 0.48$~dex more metal-rich than the mean cluster metallicity, deviating from it by $\sim 2\sigma$. Given their large individual uncertainties and the fact that their radial velocities are fully consistent with the mean cluster RV, we retain them as members, although we regard them with caution. Finally, this spectroscopic [Fe/H] was adopted to fix the metallicity in the age determination, from which we confirm a cluster age of $\sim 5$~Gyr, even if we want to emphasize that deeper photometry is needed to better characterize the main sequence and its turn-off.

\begin{table}[htpb]
\centering
\renewcommand{\arraystretch}{1.1}
\caption{Atmospheric parameters and metallicity. We list the adopted atmospheric parameters ($T_{\rm eff}$, $\log g$) and the derived [Fe/H] for each star observed with KMOS. The two stars marked with an asterisk (*) are $\sim 0.48$~dex more metal-rich than the mean cluster metallicity, deviating from it by $\sim 2\sigma$. Nevertheless, we retain them as members, since their radial velocities are fully consistent with the mean cluster RV, as shown in Fig.~\ref{fig:Garro04_RV_FeH} and Table~\ref{table:RV}.}
\begin{tabular}{lccc}
\hline\hline
\textit{Gaia} ID & $T_{\rm eff}$ & $\log g$ & [Fe/H] \\
                 &  [K] & [cgs] & \\
\hline
5864083436772875264 (*) & 4400	& 2.0 & $-0.02\pm	0.22$ \\  
5864083810333084544 & 4500	& 3.0 & $-0.39\pm	0.12$ \\  
5864083810355782784 & 4600	& 3.0 & $-0.64\pm	0.24$ \\  
5864083814699216000 & 4600	& 2.0 & $-0.56\pm	0.24$ \\  
5864083844715395968 (*) & 4400	& 2.0 & $-0.03\pm	0.21$ \\  
5864083844715404928 & 4700	& 2.0 & $-0.48\pm	0.21$ \\  
5864083879053770240 & 4700	& 2.0 & $-0.56\pm	0.26$ \\  
5864083879053774080 & 4500	& 2.5 & $-0.58\pm	0.25$ \\  
5864083879053781504 & 4600	& 2.5 & $-0.59\pm	0.23$ \\  
5864083879062387328 & 4500	& 2.5 & $-0.52\pm	0.27$ \\  
5864083879082757760 & 4600	& 2.0 & $-0.43\pm	0.22$ \\  
5864083883418740352 & 4100	& 3.0 & $-0.56\pm	0.29$ \\  
5864084050852466688 & 4700	& 2.0 & $-0.65\pm	0.24$ \\  
5864084055217451136 & 4500	& 2.0 & $-0.37\pm	0.23$ \\  
5864084055248141440 & 4800	& 2.0 & $-0.62\pm	0.25$ \\  
5864084085212562176 & 4600	& 2.5 & $-0.45\pm	0.21$ \\  
5864084089577238272 & 4800	& 2.0 & $-0.42\pm	0.22$ \\  
5864084123967628928 & 4600	& 2.0 & $-0.41\pm	0.21$ \\  
5864084261375980928 & 4600	& 2.5 & $-0.52\pm	0.23$ \\ 
\hline\hline
\end{tabular}
\label{table:metallicity}
\end{table}

\section{Discussion and conclusions}
\label{sec:Sum}
We report the discovery and cluster confirmation of Garro 04, based on a combined photometric and spectroscopic analysis of VVVX, 2MASS, and {\it Gaia} data, complemented by the first KMOS observations of cluster's targets obtained within the ESO/KMOS VVVX-GalCen Spectroscopic Survey. We identified this star cluster as a clear overdensity of red giant stars. Based on our analysis, Garro 04 is located in the Galactic disk -- likely in the thin-disk component -- at a heliocentric distance of 10.07 kpc and a Galactocentric distance of 8.3 kpc. Its derived properties suggest that Garro~04 is either an old OC or an intermediate-age cluster, with an estimated age of $5\pm 2$ Gyr and a global metallicity of $\rm{[M/H]}\approx -0.4$ dex (equivalent to the photometric metallicity [Fe/H]$=-0.55 \pm 0.2$, assuming $[\alpha/Fe] = +0.2$). This value is also in excellent agreement with the value derived star-by-star from the KMOS targets, which yields a mean cluster metallicity of $\mathrm{[Fe/H]}_{\rm spec} = -0.52 \pm 0.20$. Its orbital parameters further support this interpretation, since the cluster moves on a nearly circular, prograde orbit, trapped in disk-like motion, with an apogalactocentric distance of 9.4 kpc and a perigalactocentric distance of 7.6 kpc, very close to the Sun's orbit.  We will dedicate a follow-up paper to perform the chemical analysis and better disentangle the real nature of this cluster, and classify it as
GC, OC, or intermediate-class object. In this context, this further discovery suggests that many other small, low-luminosity clusters
are still missing from our compilation, both in the Galactic bulge and disk, as demonstrated by \cite{2024A&A...687A.214G}. Additionally -- with the caveat that a full characterization of this cluster is still required -- it is interesting to note that this
cluster might join a growing population of objects that appear too old to fit within the classical OC category, yet too young to be considered
typical GCs -- as found in the case of Patchick~126 \citep{2026A&A...706A.338G} or similarly to the old, metal-rich open cluster or migrated bulge GC NGC~6791 \citep{2018ApJ...867...34V}, to transition objects such as Palomar~1 \citep{2011ApJ...740..106S} and BH~176 \citep{2011A&A...528A..70D}, and to other intermediate-age systems recently uncovered towards
the inner Galaxy \citep{garro20, 2024A&A...687A.214G}. These systems raise the intriguing possibility that they represent an intermediate class of stellar clusters bridging the canonical populations, or alternatively, that they are the remnants of accreted systems \citep{2024A&A...691A.226C}.
Finally, we have -- for the first time -- a unique combination of spectroscopic and photometric analysis that can help to homogeneously disentangle the true nature of many of them. Our main long-term goal will be twofold: first, to confirm the existence of each cluster, as done for Garro 04 in this work; and second, to develop new techniques that fully exploit the KMOS observations, complemented by planned follow-up spectroscopy, in order to fully characterize these newly discovered star clusters.
\begin{acknowledgements}  
We thank the referee for the constructive comments, which have significantly improved the clarity and presentation of our results.\\
We are grateful to the UT1 observing team at Paranal Observatory, Dr. Jesus Corral-Santana, Dr. Martina Baratella and Sam Kim, for their support at the telescope during the nights.\\
We gratefully acknowledge the use of data from the ESO Public Survey program IDs 179.B-2002 and 198.B-2004, obtained with the VISTA telescope, as well as data products provided by the Cambridge Astronomical Survey Unit (CASU), the VISTA Science Archive (VSA), and the ESO Science Archive. The VVV and VVVX data are available in the ESO Science Archive under the following data collections and DOIs: \url{https://doi.eso.org/10.18727/archive/67} and \url{https://doi.eso.org/10.18727/archive/68}. We also acknowledge observations obtained at ESO Paranal Observatory as part of the KMOS VVVX-GalCen Spectroscopic Public Survey (ESO Programme ID 116.29EF), whose data are available through the ESO Science Archive Facility under DOI:  \url{https://doi.eso.org/10.18727/archive/38}.\\
E.R.G. gratefully acknowledges the ESO Fellowship program. I.P. acknowledges support from ANID BECAS/DOCTORADO NACIONAL 21230761.  B.P.L.P acknowledges the São Paulo Research Foundation (FAPESP), Brazil; Process Number 2025/05050-3. B.D. acknowledges support from ANID Basal project FB210003. C.O.O. acknowledges financial support from the Postdoctoral Talent Attraction Competition for Research Centers and Institutes of Universidad Andrés Bello (UNAB) 2025 under project No. DI-07-25/ATP. S.F. acknowledges the Fondecyt Regular 1240755. M.G.N.O. thanks the European Union – NextGenerationEU, M4C2 1.2, CUP C83C25000450006 and support from the SPADE project (Space Observations of Proto-planetary Accretion Disc Evolution), funded by the Italian Ministry of University and Research (MUR) through the PRORIS programme. D.M. acknowledges support by the BASAL Center for Astrophysics and Associated Technologies (CATA) through ANID grants ACE210002 and AFB 210003, and by Project Fondecyt Regular 1220724. M.G. acknowledges support by ANID Fondecyt Regular 1240755. J.G.F-T gratefully acknowledges the support provided by ANID Fondecyt Regular No.1260371.
\end{acknowledgements}

\bibliographystyle{aa}
\bibliography{references}
\end{document}